\documentclass[5p,twocolumn]{elsarticle} %review

\usepackage[export]{adjustbox}
\usepackage{mathtools}
\usepackage{bm}
\usepackage{subcaption}
\usepackage{caption}

\usepackage{times}
\usepackage{epsfig}
\usepackage{graphicx}
\usepackage{amsmath}
\usepackage{multirow}
\usepackage{xcolor}
\usepackage{amssymb}
\usepackage[multi-part-units=single]{siunitx}
\usepackage{lineno}
\usepackage{hyperref}
\newcommand{\cmmnt}[1]{\ignorespaces}
\journal{Computerized Medical Imaging and Graphics}

\newcommand*{\myDots}{\ifmmode.\kern-0.13em.\kern-0.13em.\else.\kern-0.13em.\kern-0.13em.\fi}

\biboptions{authoryear}
\begin{document}

\begin{frontmatter}

\title{Multiple Sclerosis Lesion Activity Segmentation with Attention-Guided Two-Path CNNs}

\author[mymainaddress]{Nils Gessert\corref{cor1}}
\ead{nils.gessert@tuhh.de}

\author[mymainaddress2]{Julia~Kr\"uger}
\ead{julia.krueger@jung-diagnostics.de}

\author[mymainaddress2]{Roland~Opfer}
\ead{roland.opfer@jung-diagnostics.de}

\author[mymainaddress2]{Ann-Christin~Ostwaldt}
\ead{ann-christin.ostwaldt@jung-diagnostics.de}

\author[mymainaddress3]{Praveena~Manogaran}
\ead{praveena.manogaran@usz.ch}

\author[mymainaddress4]{Hagen H. Kitzler}
\ead{hagen.kitzler@uniklinikum-dresden.de}

\author[mymainaddress3]{Sven~Schippling}
\ead{sven.schippling@usz.ch}

\author[mymainaddress]{Alexander Schlaefer}
\ead{schlaefer@tuhh.de}

\address[mymainaddress]{Hamburg University of Technology, Institute of Medical Technology, Am Schwarzenberg-Campus 3, 21073 Hamburg, Germany}

\address[mymainaddress2]{jung diagnostics GmbH, R\"ontgenstraße 24, 22335 Hamburg, Germany}

\address[mymainaddress3]{University Hospital Zurich and University of Zurich, Department of Neurology, Frauenklinikstrasse 26, 8091 Zurich, Switzerland}

\address[mymainaddress4]{Institute of Diagnostic and Interventional Neuroradiology, University Hospital Carl Gustav Carus, Technische Universität Dresden, 01062 Dresden, Germany}

\cortext[cor1]{Corresponding author}





\begin{abstract}

Multiple sclerosis is an inflammatory autoimmune demyelinating disease that is characterized by lesions in the central nervous system. Typically, magnetic resonance imaging (MRI) is used for tracking disease progression. Automatic image processing methods can be used to segment lesions and derive quantitative lesion parameters. So far, methods have focused on lesion segmentation for individual MRI scans. However, for monitoring disease progression, \textit{lesion activity} in terms of new and enlarging lesions between two time points is a crucial biomarker. For this problem, several classic methods have been proposed, e.g., using difference volumes. Despite their success for single-volume lesion segmentation, deep learning approaches are still rare for lesion activity segmentation. In this work, convolutional neural networks (CNNs) are studied for lesion activity segmentation from two time points. For this task, CNNs are designed and evaluated that combine the information from two points in different ways. In particular, two-path architectures with attention-guided interactions are proposed that enable effective information exchange between the two time point's processing paths. It is demonstrated that deep learning-based methods outperform classic approaches and it is shown that attention-guided interactions significantly improve performance. Furthermore, the attention modules produce plausible attention maps that have a masking effect that suppresses old, irrelevant lesions. A lesion-wise false positive rate of $\SI{26.4}{\percent}$ is achieved at a true positive rate of $\SI{74.2}{\percent}$, which is not significantly different from the interrater performance.

\end{abstract}


\end{frontmatter}

%\linenumbers

\section{Introduction}

Multiple sclerosis (MS) is an inflammatory autoimmune demyelinating disease of the central nervous system, which leads to disability, mostly in young adults. MS causes demyelination around nerve cells, which leads to lesions in the central nervous system. To track disease progression in the brain, magnetic resonance imaging (MRI) is often used. The fluid attenuated inversion recovery (FLAIR) sequences show white matter lesions as high-intensity regions which allow for quantification of the disease progression \citep{rovira2015evidence}. To derive quantitative parameters like lesion number and volume, lesion segmentation is required. As manual segmentation is time-consuming and error-prone \citep{egger2017mri}, several semi- and fully-automated methods have been proposed for lesion segmentation \citep{van2001automated,shiee2010topology,schmidt2012automated,roura2015toolbox}. However, manual segmentation is still the gold standard \citep{garcia2013review}. 

Recently, deep learning-based methods such as convolutional neural networks (CNNs) have gained popularity and shown promising results for brain lesion segmentation \citep{akkus2017deep,kamnitsas2017efficient}, including approaches for MS lesions \citep{danelakis2018survey}. Here, most methods rely on encoder-decoder architectures which take (a part of) an MRI volume as the input and predict a segmentation map \citep{danelakis2018survey}. Numerous architecture variations have been proposed so far, including multi-scale \citep{brosch2016deep} and cascaded \citep{valverde2017improving} approaches. 

\begin{figure*}[h!]
	\centering
   \begin{minipage}{0.25\textwidth}
  		\centering
   		\adjincludegraphics[trim={{.45\width} {.24\height} {.35\width} {.40\height}},clip,angle=90,width=1.0\linewidth]{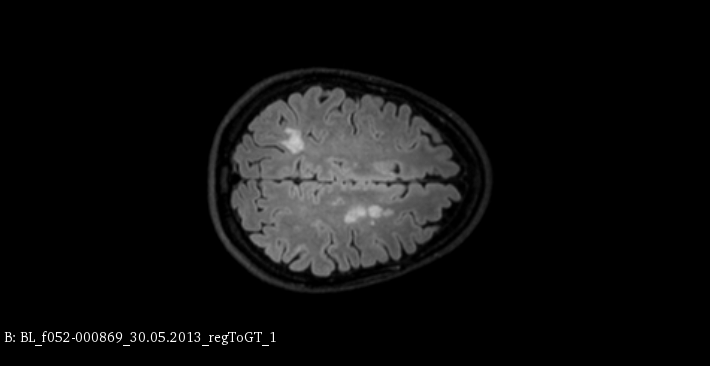}
    	\subcaption[2nd caption]{Baseline Scan}
   \end{minipage}
   %\begin{minipage}{0.195\textwidth}
  %		\centering
  % 		\adjincludegraphics[trim={{.45\width} {.24\height} {.35\width} {.40\height}},clip,angle=90,width=1.0\linewidth]{figures/BL_L_ann.png}
  %  	\subcaption[2nd caption]{Marked Baseline}
  % \end{minipage}
   \begin{minipage}{0.25\textwidth}
  		\centering
   		\adjincludegraphics[trim={{.45\width} {.24\height} {.35\width} {.40\height}},clip,angle=90,width=1.0\linewidth]{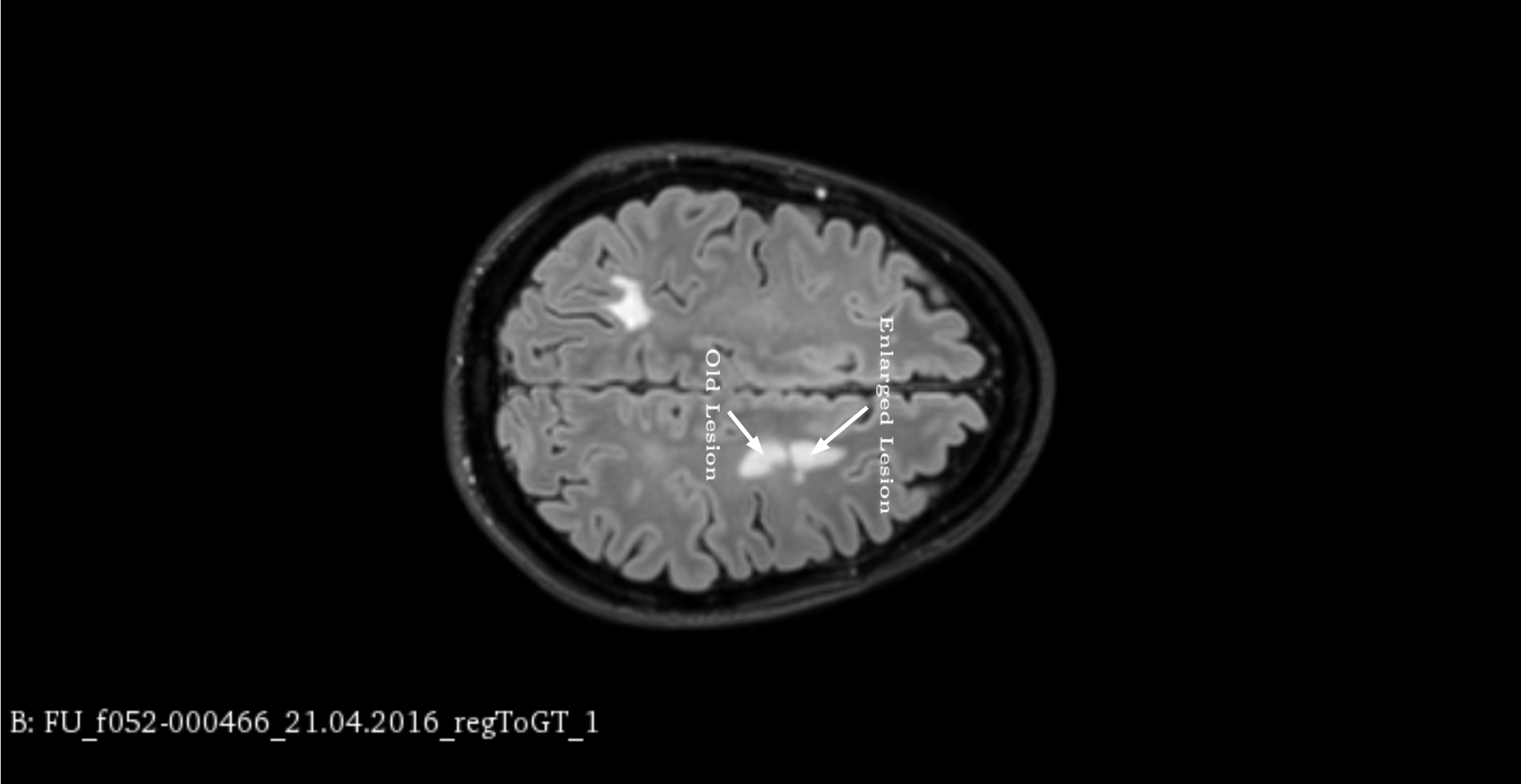}
    	\subcaption[2nd caption]{Follow-up Scan}
   \end{minipage}
   %\begin{minipage}{0.195\textwidth}
  %		\centering
  % 		\adjincludegraphics[trim={{.45\width} {.24\height} {.35\width} {.40\height}},clip,angle=90,width=1.0\linewidth]{figures/FU_ly_ann.png}
   % 	\subcaption[2nd caption]{Marked Follow-up}
   %\end{minipage}
   \begin{minipage}{0.25\textwidth}
  		\centering
   		\adjincludegraphics[trim={{.45\width} {.24\height} {.35\width} {.40\height}},clip,angle=90,width=1.0\linewidth]{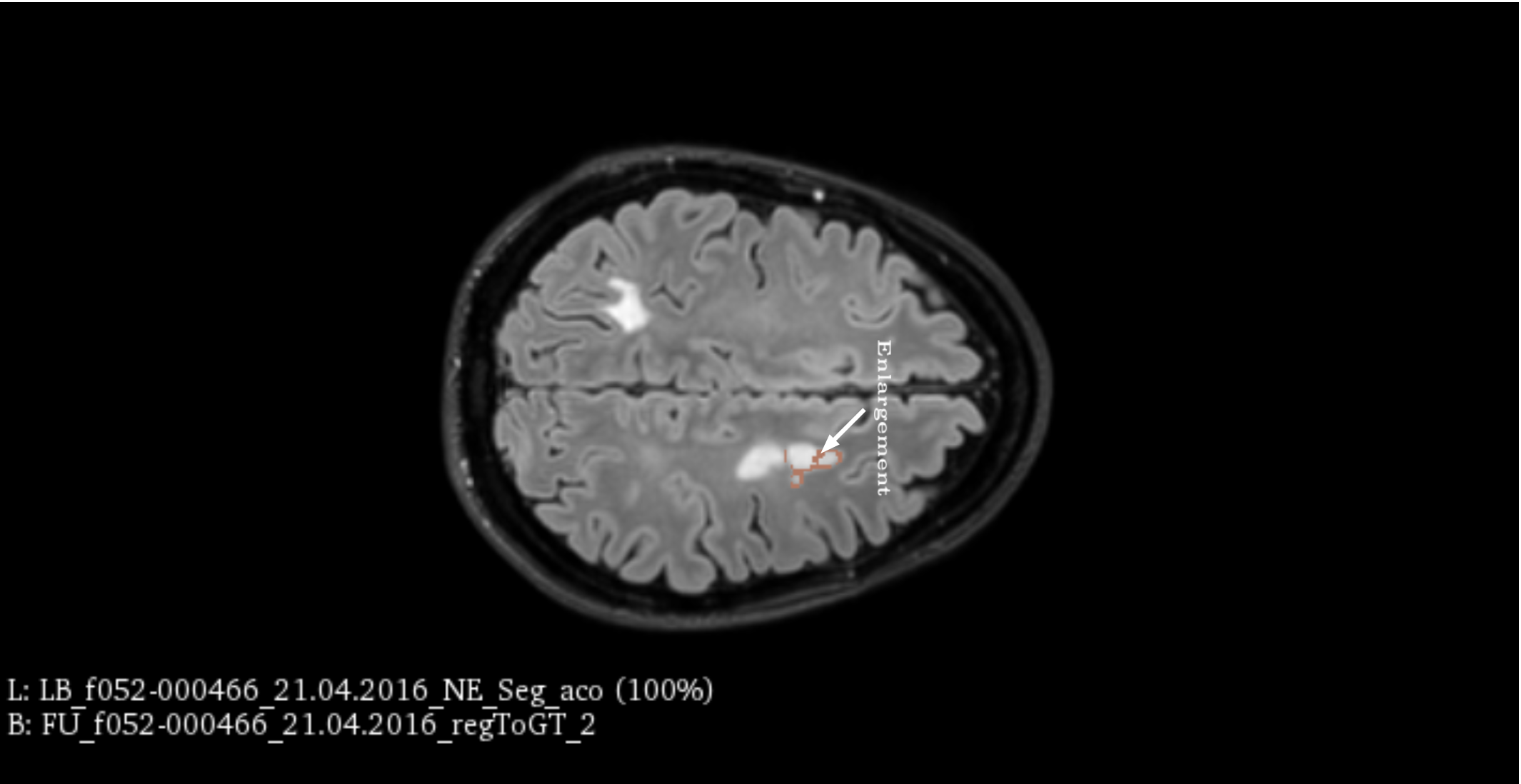}
    	\subcaption[2nd caption]{Labelled Follow-up Scan}
   \end{minipage}       
        	
	\caption{Example for lesion activity in the case of an enlarged lesion. Old lesions and old lesion material belongs the background class for task of lesion activity.}
	\label{fig:examples}
\end{figure*}

Lesion activity is defined as the appearance of new lesions and the enlargement of existing lesions \citep{mcfarland1992using}. For monitoring disease progression, lesion activity between two longitudinal MRI scans is the most important marker for inflammatory activity and disease progression in MS \citep{patti2015lesion}. Furthermore, lesion activity has been used as a secondary endpoint in numerous MS clinical drug trials \citep{sormani2013scoring,sormani2014treatment}. Thus, while extensive longitudinal analysis is not always recommended for initial diagnosis \cite{thompson2018diagnosis}, longitudinal MRI analysis for lesion activity is highly relevant in a clinical context.

 %and has been used as a secondary endpoint in numerous MS clinical drug trials.
 This problem is particularly challenging as new lesions can be small and changes are often subtle. So far, deep learning methods and most classical methods have only considered lesion segmentation for a single MRI volume. Thus, lesion activity is often derived from two independent segmentation maps which is associated with high variability and inconsistencies \citep{garcia2013review}. Therefore, other approaches made use of information from the MRI volumes instead of lesion maps only.
 For example, image differences \citep{battaglini2014automated,ganiler2014subtraction} and deformation fields \citep{cabezas2016improved,salem2018supervised} have been used to detect new lesions. Also, intensity-based approaches using local context between scans have been proposed \citep{lesjak2016validation}. Overall,  methods for detection of lesion growth have largely relied on classic image processing methods so far \citep{cheng2018multi,schmidt2019automated}.

In this paper, we address longitudinal segmentation of new and enlarged lesions using two volumes from two time points with fully-convolutional CNNs. So far, most public datasets for MS lesion segmentation provide per-scan lesion annotations without a particular focus on longitudinal lesion activity \citep{carass2017longitudinal,lesjak2018novel}. Therefore, we first create ground-truth annotations for this particular task. For each patient, we consider a baseline and a follow-up MRI scan. Three independent raters provide annotations for lesion activity.

%To combine two MRI volumes for the segmentation task, we design and evaluate different types of CNN architectures for joint processing. 
A straight-forward deep learning approach for deriving lesion activity is to use a CNN for predicting individual lesion maps for each time point and then taking the maps' difference. As previous work has demonstrated large inconsistencies for difference maps \citep{garcia2013review}, combining volumes instead of final lesion maps might be beneficial. Approaches for volume combination include taking the volume difference, volume addition or stacking volumes in the input channel dimension while using a standard single-path encoder-decoder model. However, this approach relies on high similarity between the scans and might suffer from inaccurate image registration or different acquisition parameters. Therefore, initial independent processing might be advantageous as previously shown for other deep learning problems \citep{gessert2018force}. Thus, we consider two-path encoder-decoder 3D CNNs where volumes are first processed independently by encoder paths. Then, the decoder jointly processes the combined feature maps from both volumes and predicts a segmentation of new and enlarged lesions. While initial independent processing might be beneficial \citep{gessert2018force}, we hypothesize that some degree of information exchange in the encoder paths could improve performance. Therefore, we augment the encoders by attention-guided interaction modules which allow the network to learn information exchange between the paths. We propose and evaluate different types of interaction modules at different locations inside the network.

In a preliminary abstract, we showed the feasibility of deep learning-based lesion activity segmentation \citep{krueger2019ms}. In this paper, we extend the preliminary work by two major contributions. First, we demonstrate that conventional machine learning methods for detecting lesion activity are outperformed by several deep learning-based approaches. Second, we show that attention-guided information exchange for two-path CNNs significantly improves lesion activity segmentation while producing plausible attention maps. 

\section{Methods} \label{sec:methods}

\subsection{Dataset}

\textbf{Properties and Labeling.} The dataset we use was obtained in a study that investigated MS heterogeneity at the University Hospital of Zurich, Switzerland. A 3.0T Philips Ingenia Scanner (Philips, Eindhoven, the Netherlands) was used for image acquisition using similar acquisition parameters for all scans. We use the FLAIR images as they are the recommended modality for MS lesion assessment \citep{rovira2015evidence}. For each patient, we consider a baseline scan $V_{BL}$ and a follow-up scan $V_{FU}$. The mean slice thickness is $\SI{1.2}{\milli\metre}$ and the in-plane pixel spacing is $\SI{0.92 x 0.92}{\milli\metre}$.
Three independent experts provide a set of annotations each. All experts are trained in lesion activity segmentation with several years of experience with MS cases in clinical routine. As labeling is time-consuming, we reslice $V_{FU}$ to $\SI{2}{\milli\metre}$ axial slice thickness. Then, we register $V_{BL}$ to the resliced volume using a rigid registration. The raters view both $V_{FU}$ and $V_{BL}$ simultaneously while labeling the volumes slice by slice. %Based on these annotations, we consider two label types for changed lesions. First, we only use the labels marked in $V_{FU}$ as a target (label strategy LFU). Intuitively, this tasks the models to learn \textit{which} lesion changed. Second, we set labels that overlap in $V_{FU}$ and $V_{BL}$ to the background label (label strategy LB). This resembles the learning task of \textit{how much} a lesion has changed.
Raters mark new lesions that were not present in $V_{BL}$ and also new lesion material that appeared around lesions that were already present in $V_{BL}$. Thus, we consider the task of segmenting \textit{lesion activity} which is characterized by new lesions in $V_{FU}$ and lesions that have grown between $V_{BL}$ and $V_{FU}$. Lesions are treated as enlarged if there is an increase in size by at least $\SI{50}{\percent}$, following \cite{moraal2010improved}. Old lesion regions are treated as background. Thus, the task is to learn \textit{how much} lesions have changed over time. Example images with annotations for enlarged lesions are shown in Figure~\ref{fig:examples}. We fuse the three raters' annotations by voxel-vise majority voting. %Completely new lesions that appeared in $V_{FU}$ are labelled entirely. %Note that for LB the total label volume is reduced which is important for metric interpretation.

In total, the dataset contains data from $89$ MS cases. For each case, the baseline scan and one follow-up scan are included. The mean time between BL and FU scans was $2.21 \pm 1.09$ years and patients' mean age was $36.76 \pm 8.67$ years. The labeling process resulted in \num{43} out of \num{89} cases containing new or enlarged lesions. For cases with lesion activity, on average, \num{3.52} new and enlarged lesions were present per case. We refer to the dataset as the two time point dataset 1 ($TTP_1$).

To ensure that our insights are also applicable to other datasets, we consider a second dataset for lesion activity segmentation. The dataset consists of $97$ labeled pairs from $33$ patients. All images were acquired at the Faculty of Medicine Carl Gustav Carus at Technische Universit\"at Dresden using a Siemens MAGNETOM Verio 3.0T Scanner (Siemens, Germany). The slice thickness is $\SI{1.5}{\milli\metre}$ with an in-plane pixel spacing of $\SI{1.25 x 1.25}{\milli\metre}$. The annotation strategy is the same strategy we used for dataset $TTP_1$. Annotations are provided by one rater. The mean time between BL and FU scans was $1.00 \pm 0.09$ years and patients' mean age was $39.64 \pm 10.78$ years. In total \num{41} out of \num{97} pairs contained new and enlarged lesions. On average, pairs with new and enlarged lesions contain \num{3.17} new and enlarged lesions. We refer to this dataset as the two time point dataset 2 ($TTP_2$).

We compare our approach of explicit lesion activity segmentation to the use of difference maps, derived from full lesion segmentation maps of individual scans \citep{garcia2013review}. For this purpose, we also consider a dataset for per-scan lesion segmentation, similar to a majority of the publicly available MS datasets \citep{carass2017longitudinal,lesjak2018novel}. The dataset contains \num{1500} FLAIR images, acquired during clinical routine. The images were anonymized and subsequently analyzed by jung diagnostics GmbH (Hamburg, https://www.jung-diagnostics.de/). Ground-truth lesion annotations are semi-automatically generated throughout the quality control process at jung diagnostics GmbH. Besides the different ground-truth annotation, the volumes are processed similar to the two time point dataset. We refer to this dataset as the single time point ($STP$) dataset.

\textbf{Preprocessing.} We resample $V_{BL}$ and $V_{FU}$ to $\SI{1x1x1}{\milli\metre}$ and rigidly register $V_{BL}$ to $V_{FU}$. We standardize each volume individually by subtracting the mean and dividing by the standard deviation. Then, we clip intensities at the  $1^{\textrm{st}}$ and $99^{\textrm{th}}$ percentile. We resample the ground-truth volumes to $\SI{1x1x1}{\milli\metre}$ to match the input volume size. Following Egger et al. \citep{egger2017mri}, we exclude very small lesions from our final ground-truth masks with less than $\SI{0.01}{\milli\litre}$ volume as they are not well defined and likely false positives. 

\subsection{Models}

\textbf{Classic model.} As a first reference method, we apply the longitudinal pipeline of the LST toolbox, a popular non-deep learning (classic) method for longitudinal lesion change segmentation to our data \citep{schmidt2013lst,schmidt2017bayesian}. The method first computes lesion probability maps for the individual time points using a logistic regression model. Then, a second algorithm compares the lesion probability maps and decides whether a change in lesion probability is significant or caused by FLAIR intensity variations. The method is available as a toolbox and does not require any hyperparameter tuning. %The workflow for this method is indicated in Figure~\ref{fig:lst_pipeline}.

\begin{figure}
	\centering
		\includegraphics[width=1.0\linewidth]{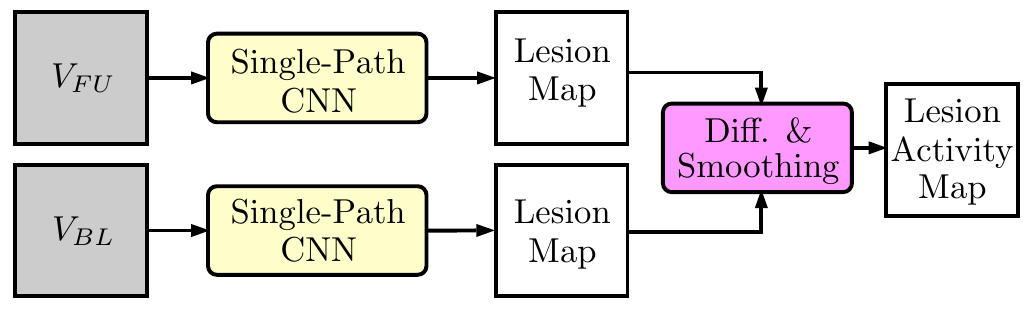}

	\caption{Prediction process for our single-path, single time point (STP) CNN approach.}
	\label{fig:stp_cnn}
\end{figure}

\textbf{Single-Path Architectures.} As a second reference, we consider a single-path (SP) fully-convolutional encoder-decoder (U-Net-like \citep{Ronneberger.2015}) architecture. After an initial convolutional layer with $32$ feature maps, we use residual blocks \citep{He.2016} both in the encoder and decoder. In the encoder, spatial feature map sizes are reduced $\num{3}$ times with convolutions having a stride of $\num{2}$ which results in $\num{4}$ spatial scales $s_i$ inside the model and a maximum reduction of the spatial size by a factor of $\num{8}$. We double the number of feature maps when the spatial size is halved. At scales $s_1$, $s_2$, $s_3$ and $s_4$ we employ $1$, $2$, $2$ and $4$ residual blocks, respectively. In the decoder, we use a single convolution followed by nearest-neighbor upsampling and a subsequent residual block at each scale. For the long-range connections between encoder and decoder we follow VoxResNet \citep{chen2018voxresnet} and use residual connections (summation) instead of feature concatenation. The model output is a dense segmentation of the same size as the input where each voxel contains the probability of having a positive label which indicates lesion activity. Due to small batch size, we use instance normalization, a variant of group normalization \citep{wu2018group}, instead of batch normalization \citep{Ioffe.2015}. 

\begin{figure}
	\centering
		\includegraphics[width=0.85\linewidth]{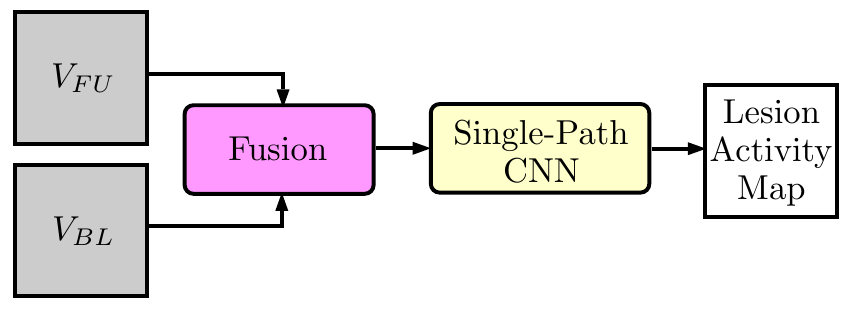}

	\caption{Prediction process for our single-path CNNs with volume fusion. Fusion is performed by addition (SP Add), difference (SP Diff) or concatenation (SP Stack).}
	\label{fig:sp_cnn}
\end{figure}

We employ this architecture for our second reference method with the STP dataset. Here, a single volume is fed into the CNN and a segmentation of all lesions is predicted. This process is repeated independently for $V_{BL}$ and $V_{FU}$. Afterward, we subtract the predicted lesion maps to obtain a map of new and enlarged lesions. Old lesions and lesions that shrunk in size are removed. We refer to this approach as STP CNN. The process for deriving lesion activity is depicted in Figure~\ref{fig:stp_cnn}.

\begin{figure*}[ht!]
	\centering
		\includegraphics[angle=90,width=0.9\textwidth]{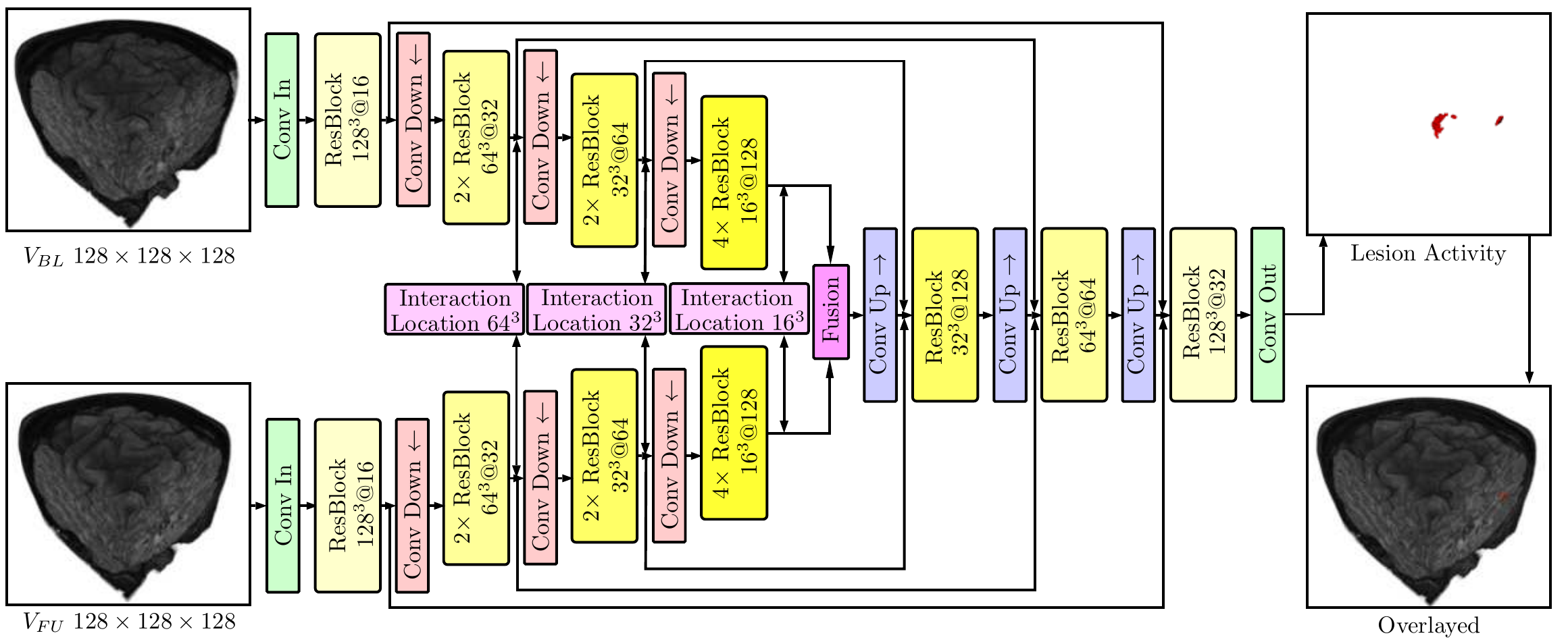}

	\caption{Our proposed two-path architecture. The output layer uses a convolution with kernel size $1\times 1\times 1$. All other kernels are of size $3\times 3\times 3$. Fusion is performed by addition (TP Add), difference (TP Diff) or concatenation (TP Stack).}
	\label{fig:model}
\end{figure*}

Also, we use this architecture for volume fusion strategies at the input where we take the volume difference (SP Diff), volume addition (SP Add) or stack the two volumes in the channel dimension (SP Stack). The key difference to the STP CNN method is that the volumes from both time points are processed jointly and lesion activity is directly predicted by the models. The prediction pipeline is shown in Figure~\ref{fig:sp_cnn}. The STP CNN architecture is the same as for the two-path architectures that we introduce next, except that a single encoder is used.

\textbf{Two-Path Architectures.} Next, we transform the single-path CNN into a two-path (TP) architecture such that the two volumes are processed in two phases. First, the volumes are processed individually in the encoder path. Second, the volumes are processed jointly in the decoder. The architecture is shown in Figure~\ref{fig:model}. Before entering the decoder, we aggregate the feature maps from both paths. We consider fusion by subtraction (TP Diff), addition (TP Add) and by feature map concatenation (TP Stack) which resembles the fusion techniques at the input for the SP CNN. Stacking the feature maps allows the network to learn which features from which part are deemed relevant. Using addition or subtraction represents voxel-wise fusion where features from both paths are directly combined. For the long-range residual connections between encoder and decoder, we concatenate the feature maps from the two encoder paths and then add the result to the decoder feature map. 

This strategy allows for individual feature learning for each time point before joint learning. To draw a connection to our single-path approaches where the volumes are processed jointly from the beginning, we also incorporate targeted information exchange between the encoder paths. For this purpose, we propose a trainable \textit{attention-guided interaction} block which controls information flow between the two paths. Inspired by squeeze-and-excitation \citep{hu2018squeeze,roy2019recalibrating} and spatial attention \citep{wang2017residual}, the attention module learns a map that suppresses spatial locations to guide the network's focus on salient regions. We consider three different attention variants at three different locations in the network, see Figure~\ref{fig:model}. 

For the first block, each path guides the other path's focus independently (attention method A). Consider features $F_{FU}^i$ and $F_{BL}^i$ of size $\mathbb{R}^{H^i \times W^i \times D^i \times C^i}$ at layer $i$ of each path. $H^i$, $W^i$ and $D^i$ are the spatial feature maps sizes and $C^i$ is the number of feature maps at layer $i$. We compute the attention map for the baseline path as $a_{BL}^i = \sigma (\mathit{conv}(F_{FU}^i))$ where $\sigma$ is a sigmoid activation function and $\mathit{conv}$ is a $1\times 1\times 1$ convolution with learnable weights $\theta_{BL}^i$. The map $a_{BL}^i$ is of size $\mathbb{R}^{H^i \times W^i \times D^i \times C^i}$ and multiplied element-wise with $F_{BL}^i$ such that $\hat{F}_{BL}^i = a_{BL}^iF_{BL}^i$. Following the concept of residual attention \citep{wang2017residual}, we finally add the modified feature map $\hat{F}_{BL}^i$ to the original feature map $F_{BL}^i$. In this way, the information in the original feature maps is preserved while the attention maps provide additional focus to relevant regions. The attention map for the follow-up path is computed symmetrically.

For the second block, we use both attention maps $F_{FU}^i$ and $F_{BL}^i$ for the computation of both attention maps (attention method B). Thus, we concatenate both feature tensors along the last dimension and compute $a_{BL}^i = \sigma (\mathit{conv}(\mathit{cat}(F_{FU}^i,F_{BL}^i)))$ with weights $\theta_{BL}^i$. The attention map has the same dimensions as the ones computed for attention method A. Similarly to A we use residual attention. We multiply the attention with the original feature map $F_{BL}^i$ and add the original feature map to the result. The map $a_{FU}^i$ is computed similarly with weights $\theta_{FU}^i$.

Third, we consider a jointly learned attention map $a_{BL}^i = a_{FU}^i$ (attention method C). We perform the same computation as for method B and share the weights $\theta_{BL}^i = \theta_{FU}^i$. In this way, method C is more efficient in terms of the number of parameters but the two paths receive attention towards the same regions. 

\begin{figure*}
	\centering
		\includegraphics[width=0.9\textwidth]{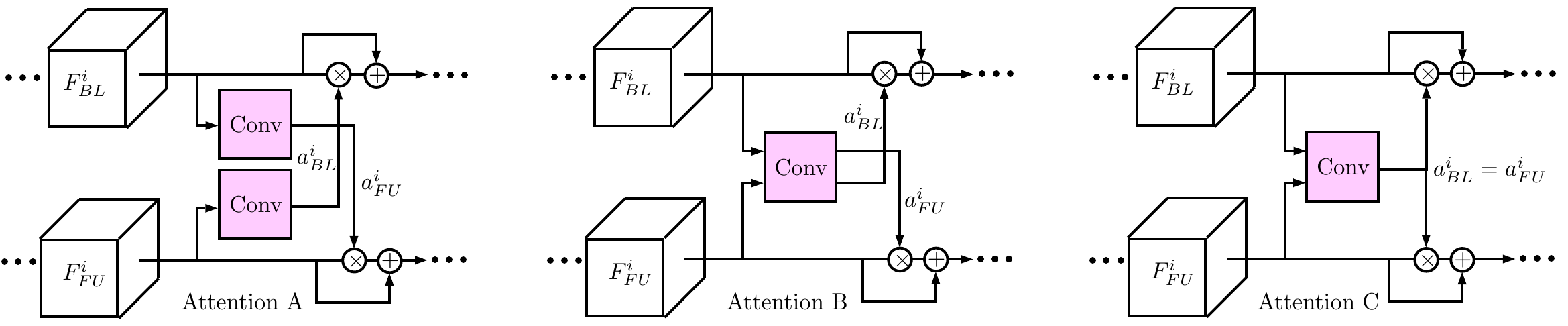}
	\caption{The three types of attention-guided interactions we employ. After the convolutions, a sigmoid activation function is applied.}
	\label{fig:attention}
\end{figure*}

\textbf{Training and Evaluation.} During training, we randomly crop subvolumes of size $128\times 128\times 128$ from the volumes. We randomly flip the volumes along the $x$, $y$ and $z$ direction with a probability of $p=0.5$. We train all models using Adam \citep{Kingma.2014} with a dice loss function and a batch size of $B=1$ for $300$ epochs. We use an initial learning rate of $l_r = \num{e-4}$ with exponential decay. For evaluation, we obtain predictions for entire volumes by taking $\num{27}$ evenly spaced, overlapping subvolumes from each volume, and obtain their corresponding predicted subvolume lesion activity map. The subvolumes are of size $128\times 128\times 128$. We merge all predicted subvolume lesion activity maps to one lesion activity map by calculating the mean probability in overlapping regions. Then, we calculate metrics using the ground-truth label volumes from each rater. The final values are averaged across all raters. The strategy is inspired by multi-crop evaluation, as employed by \cite{simonyan2014very}.

\subsection{Evaluation, Metrics and Experiments}

\textbf{Data Split.} We use a three-fold cross-validation strategy for validation and testing. We define three mutually exclusive folds with $26$ cases each. Each fold is divided into a validation and a testing split. For hyperparameter tuning, we train on the remaining cases for each fold and chose hyperparameters based on validation performance, averaged across the three validation splits. For testing, we train on all cases except for the test splits for each fold. Performance metrics are reported for the test splits. Metrics are calculated for all cases and aggregated across all test splits. The same strategy is used for datasets $TTP_1$ and $TTP_2$. For experiments with dataset $STP$, training for single time point segmentation is performed on all $STP$ cases. Testing is performed on the test splits of dataset $TTP_1$. Experiments with two path models are performed on $TTP_1$, unless indicated otherwise.

\textbf{Metrics.} In terms of metrics, we consider lesion-wise metrics and the dice coefficient. The number of new and enlarging lesions are the most relevant indicators of disease progression. Therefore, our primary metrics are the lesion-wise true positive rate (LTPR) and lesion-wise false positive rate (LFPR). We define lesions as groups of 26-connected voxels. LTPR is the number of lesions that overlap in a prediction and ground-truth map divided by the total number of lesions in the ground-truth map. LFPR is the number of lesions that do not overlap in a prediction and ground-truth map divided by the total number of predicted lesions. While LFPR and LTPR are indicators for lesion presence, we use the dice score to quantify lesion overlap. Thus, in case a predicted and a ground-truth lesion overlap, we calculate the dice score for that lesion. Note, that calculating a dice score for non-overlapping lesions is not meaningful as the score is always $0$. All three metrics depend on a decision threshold (typically \num{0.5}) for a model's predicted probabilities. For each model, we chose the optimal threshold based on ROC analysis where the sum of LTPR and $1-$LFPR is maximized. For each metric we provide the mean value as well as the $25^{\textrm{th}}$ and $75^{\textrm{th}}$ percentile range. We test for a significant difference in the median of the different metrics using Wilcoxon's signed-sum test with confidence level of $\alpha = 0.05$. Thus, we claim statistical significance if the test yields $p < 0.05$. For dataset $TTP_1$, we consider the mean interrater performance, calculated over all pairwise comparisons between the three raters for comparison to model results.

%and the number of false positives (FP).
\textbf{Experiments.} We first consider the two reference scenarios with the STP CNN and the approach using the LST toolbox. We compare this approach to several single-path, fusion-based deep learning methods including volume subtraction and addition, channel stacking and basic two-path architectures. Second, we provide results for two-path architectures, enhanced by our novel attention-guided interaction modules. We consider different attention blocks, different attention locations, and an additional evaluation on dataset $TTP_2$. Last, we present qualitative results for our attention maps.

\begin{table}[!t]
\renewcommand{\arraystretch}{1.2}
\setlength{\tabcolsep}{6pt}
\caption{Mean value and percentile ranges of dice, LTPR, and LFPR for the reference methods LST and STP CNN in comparison to single- and two-path deep learning methods with different fusion techniques (difference, addition, channel stacking). The best performing method for each metric is marked bold.}
%All metrics except for FP are given in percent.
\label{tab:res}
\centering
\begin{tabular}{l l l l l}
 & Dice & LTPR & LFPR \\
\hline
 Interrater & $67.2 (67,81)$ & $72.6 (50,100)$ & $22.9 (0,33)$  \\
 \hline
LST & $50.0 (41,65)$ & $65.8 (50,100)$ & $65.6 (60,84)$  \\
STP CNN & $52.2 (39,72)$ & $66.9 (50,100)$ & $55.1 (32,86)$ \\
\hline
SP Diff & $48.4 (37,72)$ & $52.9 (22,100)$ & $38.5 (0,58)$  \\
SP Add & $46.3 (34,70)$ & $54.0 (30,100)$ & $43.6 (0,64)$  \\
SP Stack & $56.2 (52,75)$ & $59.7 (38,100)$ & $33.6 (0,50)$  \\
\hline
TP Diff & $58.1 (50,78)$ & $61.5 (50,100)$ & $\pmb{28.3 (0,50)}$ \\
TP Add & $59.2(56,77)$ & $\pmb{63.7 (43,100)}$ & $30.4 (0,50)$  \\
TP Stack & $\pmb{58.3 (56,77)}$ & $60.6 (38,100)$ & $31.7 (0,50)$  \\
\hline
%\parbox[t]{2mm}{\multirow{8}{*}{\rotatebox[origin=c]{90}{Registered}}}
\end{tabular}
\end{table}

\section{Results}

First, we consider results for the reference approaches compared to several single-path, fusion-based deep learning methods, see Table~\ref{tab:res}. The LST method's LFPR is very high. Note that the LST result is not directly comparable to the other result as the predictions cannot be evaluated at the respective interrater LFPR. Comparing the STP CNN to the SP CNNs, metrics are similar for addition and subtraction of baseline and follow-up scan. However, the performance difference is statistically significant between SP Stack and STP CNN both for the LTPR and LFPR. For the TP CNNs, it is notable that all variants perform better than the SP CNN approaches. In particular, TP Diff and TP Add significantly outperform all other SP variants in terms of the LTPR and LFPR. There is no significant difference in performance between the three variants. Notably, the performance difference between all models and the interrater performance is statistically significant for the LTPR and LFPR. 

\begin{table}[!t]
\renewcommand{\arraystretch}{1.2}
\setlength{\tabcolsep}{6pt}
\caption{Mean value and percentile ranges of dice, LTPR, and LFPR in percent for our different attention methods A, B and C at location $16^3$. The best performing attention method for each two-path fusion type (difference, addition or channel stacking) is marked bold.}
\label{tab:res_att}
\centering
\begin{tabular}{l l l l}
 & Dice & LTPR & LFPR \\
\hline 
 Interrater & $67.2 (67,81)$ & $72.6 (50,100)$ & $22.9 (0,33)$  \\
 \hline
TP Diff & $58.1 (50,78)$ & $61.5 (50,100)$ & $28.3 (0,50)$  \\
TP Diff A & $62.4 (57,78)$ & $72.0 (50,100)$ & $27.6 (0,50)$  \\
TB Diff B & $61.9 (55,79)$ & $72.4 (50,100)$ & $28.0 (0,50)$ \\
TP Diff C & $\pmb{63.2 (58,79)}$ & $\pmb{73.2 (50,100)}$ & $\pmb{26.2 (0,50)}$ \\
\hline
TP Add & $59.2(56,77)$ & $63.7 (43,100)$ & $30.4 (0,50)$  \\
TP Add A & $60.9 (53,79)$ & $69.9 (50,100)$ & $31.4 (0,50)$ \\
TB Add B & $61.0 (58,77)$ & $69.4 (50,100)$ & $28.9 (0,50)$  \\
TP Add C & $\pmb{62.2 (54,76)}$ & $\pmb{74.2 (50,100)}$ & $\pmb{26.4 (0,50)}$  \\
\hline
TP Stack & $58.3 (56,77)$ & $60.6 (38,100)$ & $31.7 (0,50)$  \\
TP Stack A & $64.7 (59,81))$ & $71.2 (50,100)$ & $28.5 (0,50)$  \\
TP Stack B & $63.2 (60,79)$ & $70.9 (50,100)$ & $27.5 (0,50)$ \\
TP Stack C & $\pmb{65.6 (64,79)}$ & $\pmb{73.1 (56,100)}$ & $\pmb{26.9 (0,50)}$  \\
\hline
%\parbox[t]{2mm}{\multirow{8}{*}{\rotatebox[origin=c]{90}{Registered}}}
\end{tabular}
\end{table}

Second, we present results for our different attention mechanisms at location $16^3$ for the three two-path models TP Diff, TP Add, and TP Stack, see Table~\ref{tab:res_att}. For all three fusion methods, the attention blocks improve performance. Comparing the attention methods to their baseline without attention, the median LTPR and dice coefficients are significantly different across all fusion and attention methods. For the LFPR, the attention methods also improve performance but the difference in the median is not significant. Comparing attention method C to the interrater performance, there is no significant difference in the median of all metrics for all fusion techniques. 

\begin{table}[!t]
\renewcommand{\arraystretch}{1.2}
\setlength{\tabcolsep}{6pt}
\caption{Mean value and percentile ranges of dice, LTPR, and LFPR in percent for different attention locations with our three two-path models with different fusion techniques (difference, addition and channel stacking) and attention C. The location is indicated by the size of the feature maps at that level. \textit{All} refers to attention modules at all three locations.}
\label{tab:res_loc}
\centering
\begin{tabular}{l l l l}

 & Dice & LTPR & LFPR \\
\hline
 Interrater & $67.2 (67,81)$ & $72.6 (50,100)$ & $22.9 (0,33)$ \\
 \hline
TP Diff $16^3$ & $63.2 (58,79)$ & $\pmb{73.2 (50,100)}$ & $\pmb{26.2 (0,50)}$  \\
TP Diff $32^3$ & $60.1 (54,76)$ & $72.2 (50,100)$ & $28.5 (0,50)$ \\
TP Diff $64^3$ & $62.0 (60,77)$ & $70.0 (50,100)$ & $29.1 (0,50)$  \\
TP Diff all & $\pmb{64.2 (59,77)}$ & $73.0 (50,100)$ & $27.8 (0,50)$  \\ \hline
TP Add $16^3$ & $62.2 (54,76)$ & $\pmb{74.2 (50,100)}$ & $\pmb{26.4 (0,50)}$  \\
TP Add $32^3$ & $61.0 (58,78)$ & $70.2 (50,100)$ & $29.8 (0,50)$  \\
TP Add $64^3$ & $62.8 (59,79)$ & $71.4 (50,100)$ & $30.0 (0,50)$ \\ 
TP Add all & $\pmb{64.2 (63,79)}$ & $71.9 (50,100)$ & $27.8 (0,50)$ \\
\hline
TP Stack $16^3$ & $\pmb{65.6 (64,79)}$ & $\pmb{73.1 (56,100)}$ & $\pmb{26.9 (0,50)}$  \\
TP Stack $32^3$ & $62.4 (55,79)$ & $72.2 (50,100)$ & $28.4 (0,50)$  \\
TP Stack $64^3$ & $62.8 (53,79)$ & $72.0 (50,100)$ & $29.0 (00,50)$  \\
TP Stack all & $62.5 (55,77)$ & $72.5 (50,100)$ & $29.3 (0,50)$  \\
\hline
%\parbox[t]{2mm}{\multirow{8}{*}{\rotatebox[origin=c]{90}{Registered}}}
\end{tabular}
\end{table}

\begin{figure*}
\begin{center}
   %\begin{subfigure}{0.245\linewidth}
   \begin{minipage}{0.33\textwidth}
  		\centering
   		\includegraphics[width=1.0\linewidth]{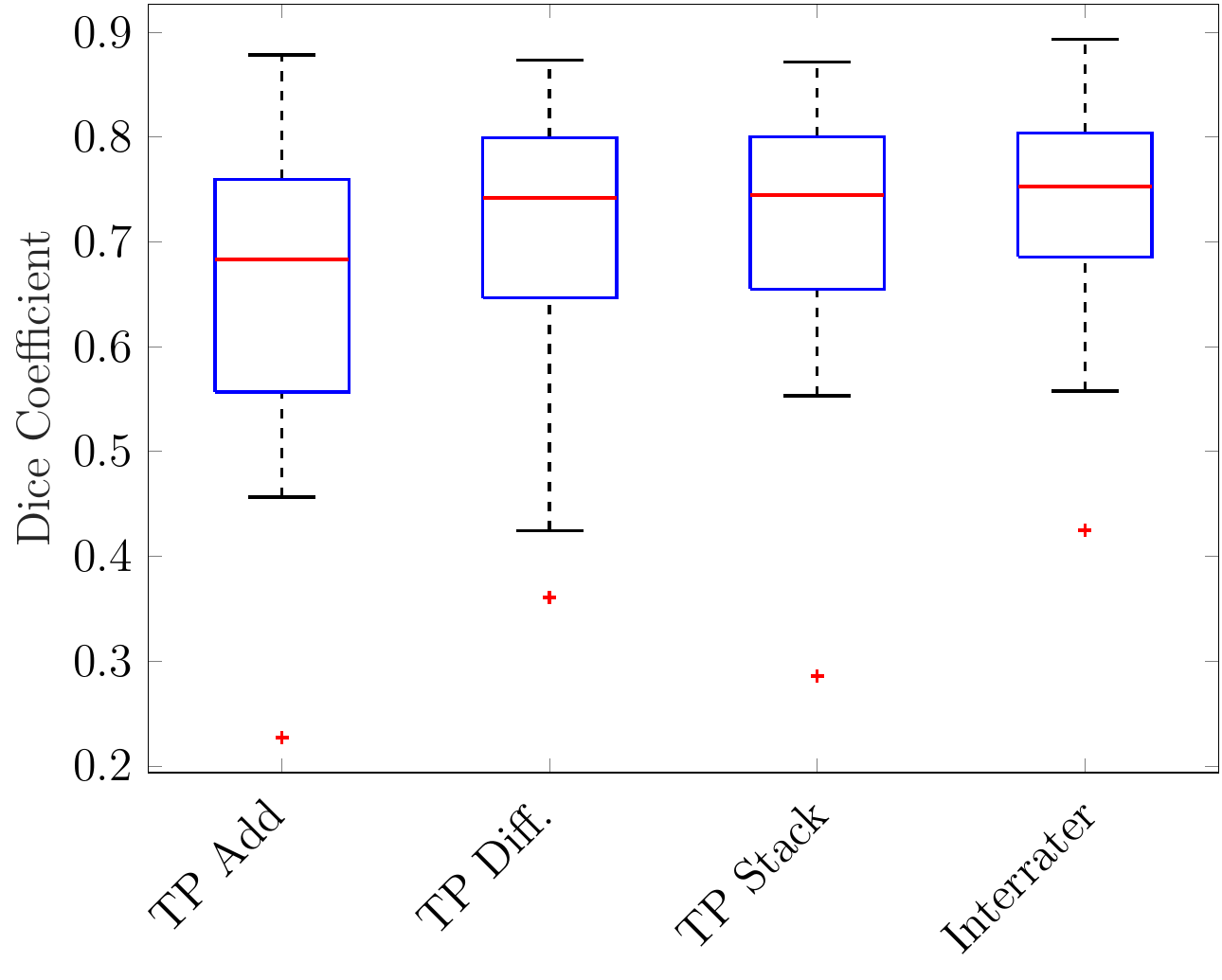}
    	%\subcaption[2nd caption]{Dice ($\textrm{LFPR}=\SI{22.9}{\percent}$)}
   \end{minipage}
   %\end{subfigure}
   %\begin{subfigure}{0.245\linewidth}   
   \begin{minipage}{0.33\textwidth}
  		\centering   
   		\includegraphics[width=1.0\linewidth]{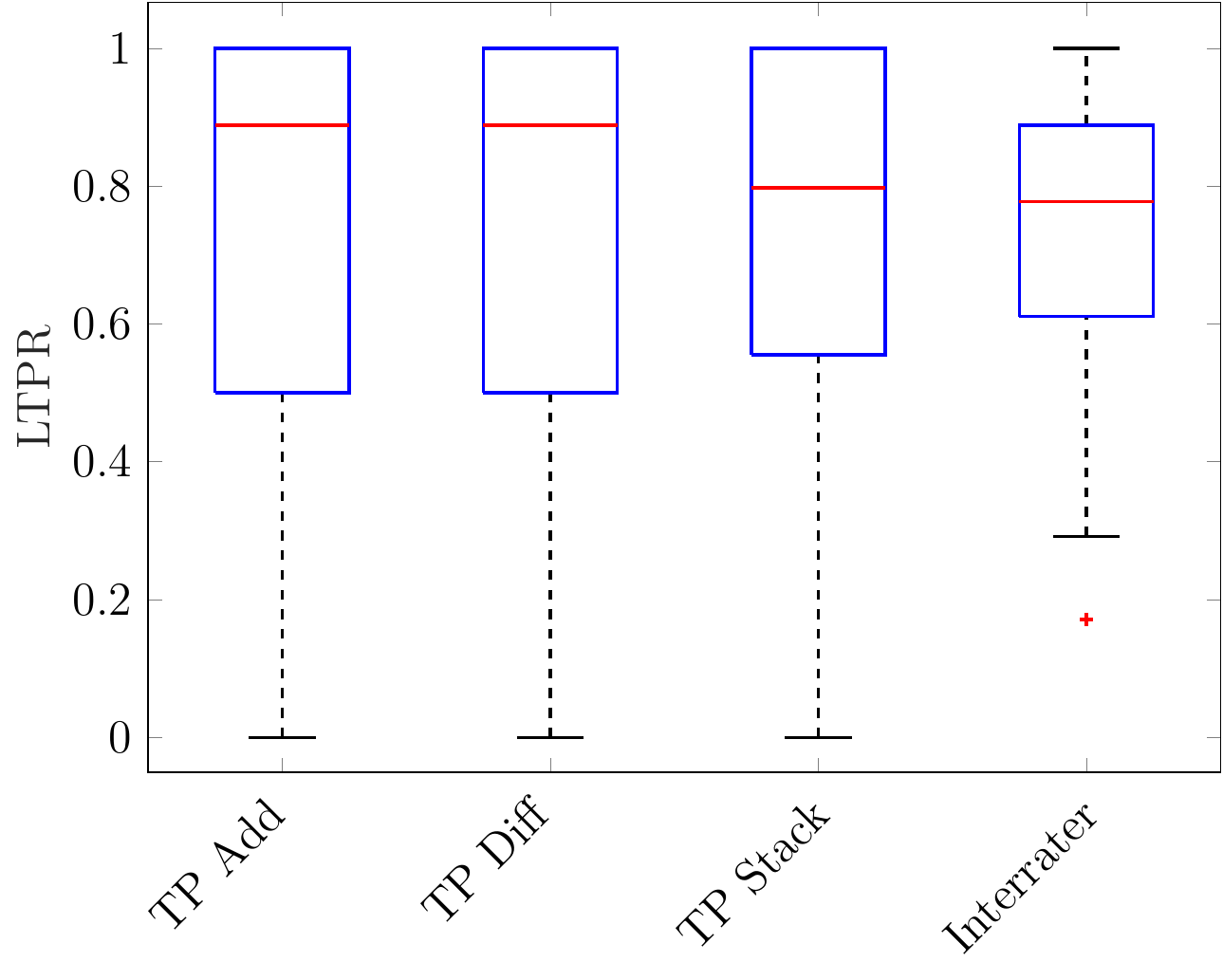}
    	%\subcaption[2nd caption]{LTPR ($\textrm{LFPR}=\SI{22.9}{\percent}$)}
   \end{minipage}   		
   %		\caption{convGRU-ResNet3D DS 1}
   %\end{subfigure}   
   %\begin{subfigure}{0.245\linewidth}   
   \begin{minipage}{0.33\textwidth}
  		\centering      
   		\includegraphics[width=1.0\linewidth]{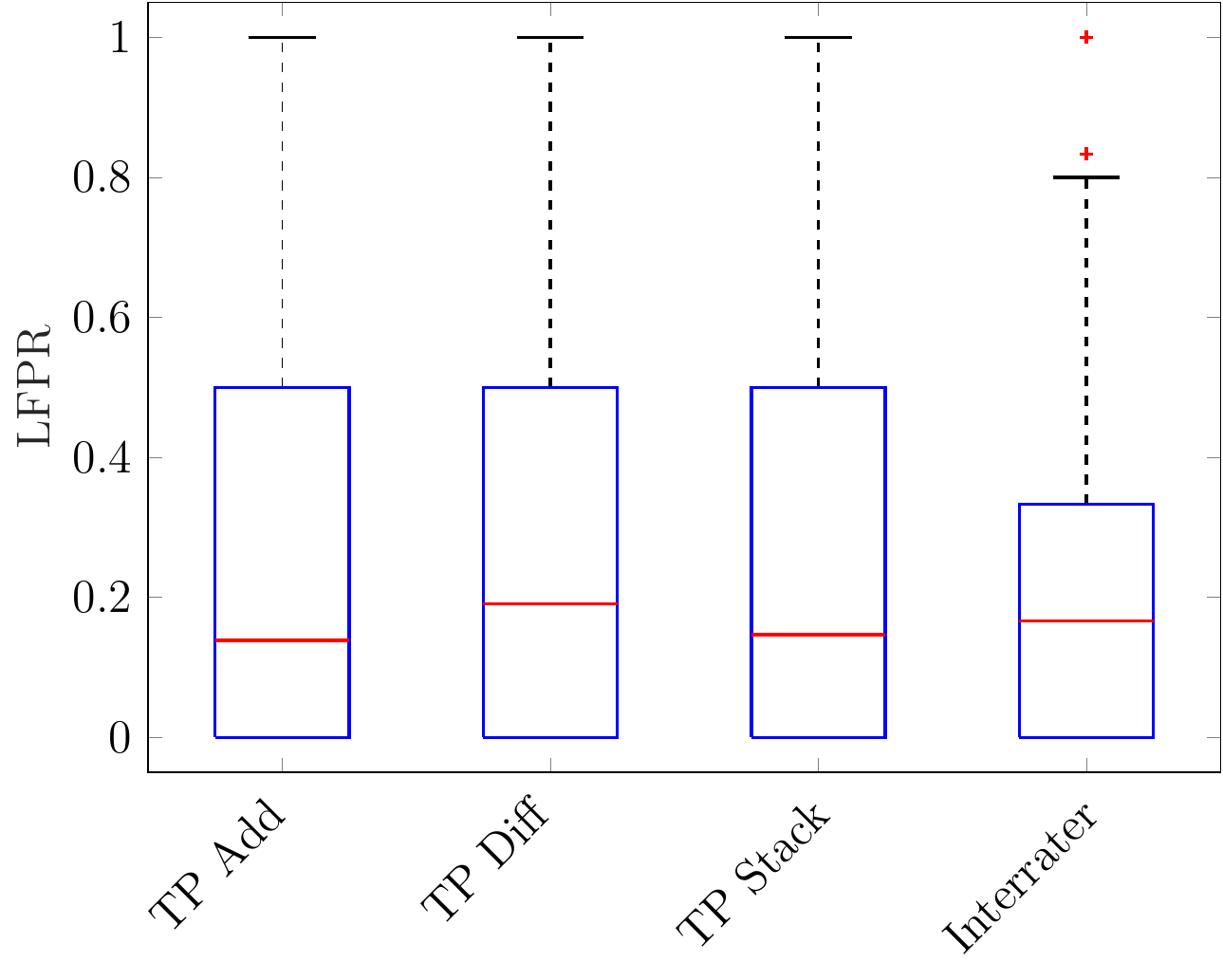}
    	%\subcaption[2nd caption]{Dice ($\textrm{LTPR}=\SI{72.6}{\percent}$)}
   \end{minipage}     		
  		
\end{center}
   \caption{Boxplots for the three two-path models with attention (TP C) at location $16^3$ compared to the interrater performance.}
   %For each configuration, a new model is trained.
\label{fig:boxplots}
\end{figure*}

Third, we demonstrate how the location of the attention block affects performance for the two-path models, see Table~\ref{tab:res_loc}. Attention blocks at location $16^3$ tend to perform best. When attention blocks are placed further towards the model input, performance tends to go down. When placing attention blocks at all three locations at the same time, LTPR and LFPR do not improve further. The dice score, however, improves for some scenarios.

Fourth, we show boxplots of the three TP architectures with attention C at location $16^3$, see Figure~\ref{fig:boxplots}. In terms of the LFPR, the distribution of each model is similar to the interrater distribution. For the LTPR the same observation can be made. The distribution of the interrater dice scores is slightly higher than the dice score of TP Add C. For TP Diff C and TP Stack C, the distribution has a similar median and there is no significant difference in the median, compared to the interrater performance.

\begin{table}[!t]
\renewcommand{\arraystretch}{1.2}
\setlength{\tabcolsep}{6pt}
\caption{Mean value and percentile ranges of dice, LTPR, and LFPR in percent for our different attention methods A, B and C at location $16^3$ for the second dataset $TTP_2$. The best performing attention method for each two-path fusion type (difference, addition or channel stacking) is marked bold.}
\label{tab:res_att_dd}
\centering
\begin{tabular}{l l l l}
 & Dice & LTPR & LFPR \\
 \hline
TP Diff & $60.6 (56,79)$ & $67.5 (50,100)$ & $27.0 (00,43)$ \\
TP Diff A & $61.9 (56,79)$ & $75.1 (50,100)$ & $\pmb{24.9 (00,38)}$  \\
TB Diff B & $63.1 (56,81)$ & $75.3 (67,100)$ & $28.2 (00,50)$  \\
TP Diff C & $\pmb{64.1 (62,80)}$ & $\pmb{76.5 (67,100)}$ & $25.0 (00,40)$ \\
\hline
TP Add & $60.9 (54,79)$ & $68.6 (50,100)$ & $34.2 (00,54)$  \\
TP Add A & $\pmb{63.3 (61,83)}$ & $71.6 (50,100)$ & $28.6 (00,50)$  \\
TB Add B & $60.9 (53,75)$ & $\pmb{75.3 (67,100)}$ & $\pmb{22.7 (00,33)}$  \\
TP Add C & $61.2 (58,79)$ & $73.7 (56,100)$ & $26.4 (00,50)$  \\
\hline
TP Stack & $62.7 (54,79)$ & $71.8 (50,100)$ & $28.1 (00,50)$  \\
TP Stack A & $\pmb{65.2 (57,81)}$ & $\pmb{79.4 (67,100)}$ & $22.8 (00,33)$  \\
TP Stack B & $64.6 (61,79)$ & $76.0 (67,100)$ & $26.5 (00,50)$ \\
TP Stack C & $64.3 (55,81)$ & $77.2 (67,100)$ & $\pmb{21.8 (00,33)}$  \\
\hline
%\parbox[t]{2mm}{\multirow{8}{*}{\rotatebox[origin=c]{90}{Registered}}}
\end{tabular}
\end{table}

Fifth, we investigate whether our attention method is also advantageous on another dataset, using dataset $TTP_2$. The results for attention location $16^3$ are shown in Table~\ref{tab:res_att_dd}. In general, the attention methods improve performance once again, across multiple fusion strategies. The difference is largest for the LTPR where all attention methods across all fusion strategies significantly outperform the baseline. For the other metrics, there is also consistent improvement.

\begin{figure*}
	\centering
	\includegraphics[width=1.0\textwidth]{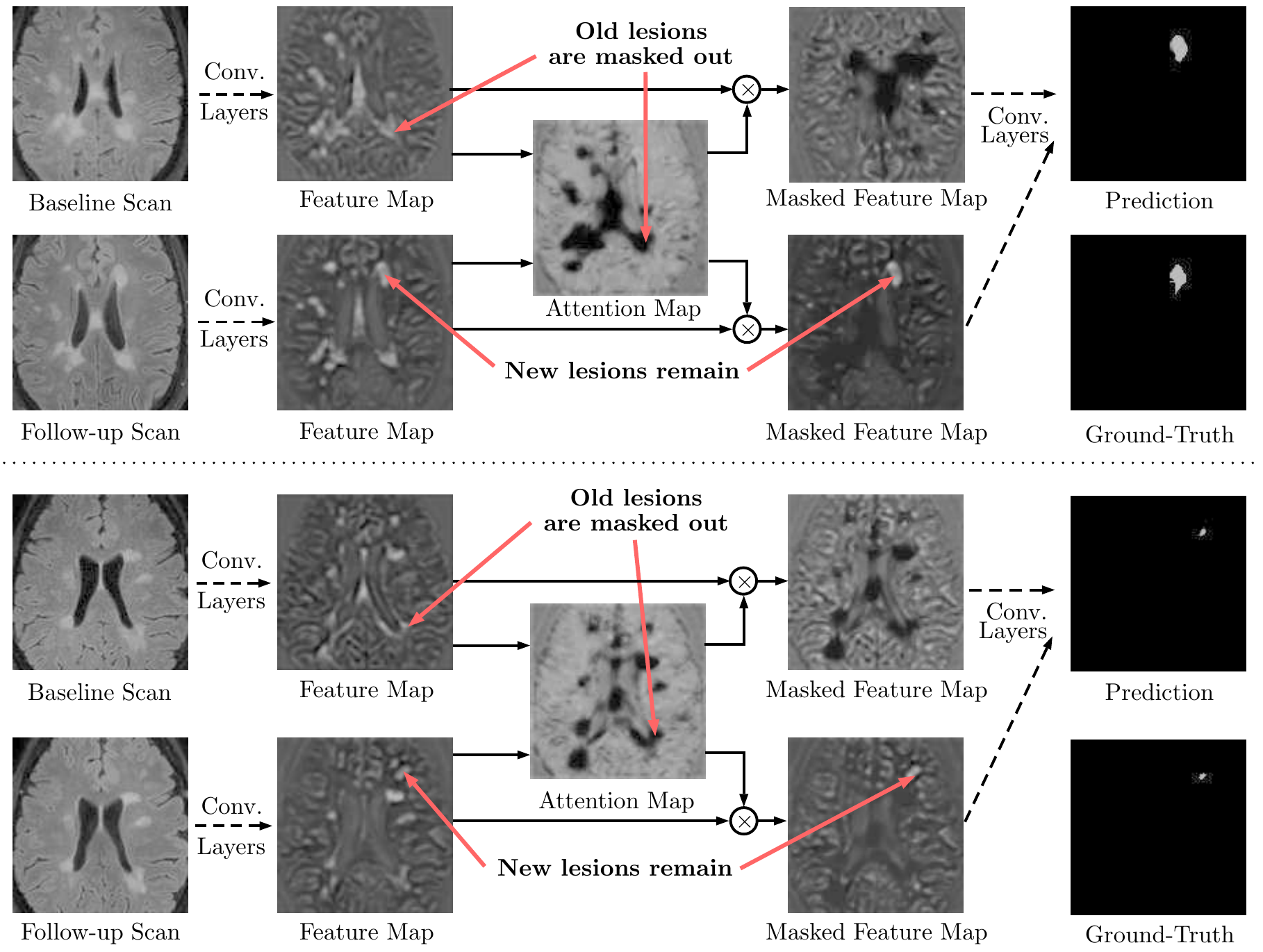}
	\caption{Visualization of the effect of our attention method for the model TP Stack C $64^3$. Two example cases are shown (top and bottom). Left, an axial slice of the baseline and follow-up scan are shown. In the center, example features maps, the attention maps and the masked feature maps are shown. Right, the final model prediction and ground-truth are shown.}
	\label{fig:qualitative}
\end{figure*}

Last, we show qualitative results for our attention mechanism, see Figure~\ref{fig:qualitative}. The Figure shows two examples for the effect of our attention mechanism for the model TP Stack C $64^3$. Feature maps from the baseline and follow-up scan are used to compute an attention map. This map is then multiplied with each feature map. The attention maps show very low-intensity regions where old lesions are present in the baseline scan. In the follow-up scan's feature map, regions corresponding to lesions show a high intensity. After applying the attention maps, a masking effect can be observed.  Lesions that were already present in the baseline scan are masked out by the attention map. Thus, in the follow-up scan's feature map, only small high-intensity regions remain. Comparing these regions to the final predictions and ground-truth maps shows that the remaining high-intensity regions correspond to a new lesion. Therefore, the attention maps masked out old lesions while preserving the new lesions.

\section{Discussion}
%So far, lesion segmentation from single MRI scans has been approached with the use of CNNs. 
We address the problem of multiple sclerosis lesion activity segmentation from two time points using deep learning models. Compared to lesion segmentation in a single scan, the task of lesion activity segmentation is much more challenging as lesion size can be small and it is difficult to distinguish lesion activity from slight registration errors and intensity variations between two scans. As most MS lesion segmentation datasets are created for single-scan lesion segmentation \citep{carass2017longitudinal,lesjak2018novel}, we created two datasets with ground-truth annotations for new and enlarged lesions. 

As a reference for lesion activity segmentation from two MRI scans, we consider two methods. The first method is the LST toolbox, an established classic, non-deep learning method \citep{schmidt2019automated} that uses FLAIR images for lesion prediction while the second method uses the typical approach of individual lesion map differences \citep{garcia2013review}. We compare these references to deep learning-based, single-path fusion techniques for baseline and follow-up scan aggregation by addition and difference. This follows the rationale of previous methods that relied on difference images \citep{ganiler2014subtraction}. The two reference methods perform similar to these approaches which indicates that more advanced architectures are required to improve lesion activity segmentation. When stacking baseline and follow-up scan into the input channel dimension, which is typically used for color channels in the natural image domain, performance increases significantly. This indicates that individual volume processing is advantageous for this task. Therefore, we design a two-path architecture where volumes are processed individually in the encoder path. The two paths are combined with three different fusion methods which all perform better than the single-path models with volume fusion and the two reference methods. In particular, using addition or difference for fusion leads to statistically significant differences in terms of our primary metrics, the LTPR and LFPR. However, all these methods show a performance that is still significantly different from the interrater performance. 

To improve performance further, we therefore propose attention-guided interaction modules. These modules allow for information exchange between the two encoder paths before feature fusion at the model's decoder. Overall, we can observe a performance improvement across all fusion techniques and attention modules for the LFPR and the LTPR. In particular, attention methods C stands out as it performs best and the performance improvement over the normal two-path models is statistically significant for the LTPR across all fusion techniques while maintaining a low LFPR. For all metrics, there is no significant difference between the interrater performance and our two-path model with attention method C. This suggests the usability of our proposed method for clinical practice. Given that our method performs similar to human raters, its application could significantly reduce the workload for radiologists by providing usable, high-quality lesion activity segmentation maps. 

This insight is confirmed on a second, independent dataset that was acquired at a different location with a different device. Here, it is notable that the other attention methods A and B also lead to a high performance. Furthermore, we study how the location of our attention blocks inside the network affects performance, see Table~\ref{tab:res_loc}. Here, we can observe a decrease in performance if the modules are moved further away from the fusion point towards the model input. This could indicate that our attention blocks are mostly effective for guiding feature fusion. If the information exchange happens too early inside the network, the effect potentially decreases until the actual feature fusion occurs. In addition, we also consider a scenario where we place attention blocks at all three locations. However, we do not observe a synergy effect as the LTPR and LFPR do not improve further. This indicates that an additional flow of information between path does not provide an additional benefit and that a single attention module is likely sufficient to guide information exchange.
% \citep{fahrbach2013relating} for monitoring clinical disease activity.

Last, we provide a qualitative evaluation of our attention method. While quantitative results already imply the effectiveness of the proposed attention modules, we also provide a visualization of the computed attention maps, see Figure~\ref{fig:qualitative}. The attention maps appear to provide an effect that is similar to masking. This is particularly visible for the feature maps derived from the follow-up scan. After the masking process, the old lesions are attenuated and only a new lesion remains with a high intensity. This implies that our module design for information exchange between the two paths is very effective as key information is successfully transferred. Using attention maps could be particularly interesting in relation to biomarker discovery. While deep learning methods are often regarded as "black boxes," due to the difficult interpretability of features being learned, attention methods might be able to reveal the inner workings of CNNs. Thus, for future work, attention methods could be analyzed with respect to their suitability of highlighting new features or criteria that were learned by CNNs that localize MS lesion activity.

Following \cite{rovira2015evidence}, we use FLAIR images in our study. In the future, the datasets could be extended by other MR modalities that have also been studied in the context of MS lesions \citep{van1998histopathologic}. As deep learning methods have been employed with different MR imaging modalities \cite{valverde2017improving}, our models could also be directly trained with other MR imaging modalities, for example, using channel stacking. Furthermore, our datasets could be extended by including annotations for disappearing lesions between the baseline and the follow-up scan. While lesion activity, the appearance of new and enlarging lesions, is of primary interest for quantifying disease progression \cite{patti2015lesion}, studying the disappearance of MS lesions could also be of interest.
%We leverage the concept of residual attention \citep{wang2017residual}. This ensures that the modules do not deteriorate already computed features as the original features maps are always preserved, see Figure~\ref{fig:attention}. If necessary, the model could pull the attention module's contribution to zero which would result in the normal two-path models without attention modules.

As a result, we first find that individual processing of baseline and follow-up scan with two-path architectures is more effective than using the classic LST method, lesion map differences or single-path CNNs with initial volume fusion. Then, we successfully recover information exchange between baseline and follow-up scan with our proposed attention modules which show promising results, both quantitatively and qualitatively. Our method is designed for the analysis of two MRI scans from two time points, which matches the clinical workflow of practitioners for determining lesion activity. Often, there are more time points available for long-term patients. While our method's application to multiple time points is straight forward with multiple pairwise comparisons, future work could explore methods for the simultaneous analysis of several time points. Another important step would be to incorporate the problem of multiple MRI scanners. In clinical practice, scans will often be acquired with different scanners or scanning parameters. One way to achieve robustness towards the intensity variations caused by different scanners would be to use a large multi-scanner dataset. As large datasets are often difficult to obtain, another promising approach could be deep learning-based domain adaptation \cite{ganin2015unsupervised} for achieving invariance towards different scanner properties. In particular, adversarial domain adaptation strategies \cite{kamnitsas2017unsupervised} could be added as an augmentation to our models in future work.

\section{Conclusion}

We address the task of segmenting new and enlarging multiple sclerosis lesions that occurred between a baseline and a follow-up MRI scan. So far, deep learning methods have hardly been used for this problem despite their success in related tasks. Therefore, we investigate different deep learning approaches for lesion activity segmentation. We find that two-path convolutional neural networks outperform both a classic reference and single-path models. Based on this observation we propose attention-guided interactions that improve two-path models further by providing information exchange between processing paths. We demonstrate that our method significantly outperforms several other models and that our method's performance is close to the human interrater level. For future work, our method could be applied to other problems related to tracking disease progression over time.

\section*{Acknowledgment}

This work was partially supported by AiF grant number ZF4268403TS9.

\section*{References}

\bibliography{egbib}

\end{document}